\documentstyle[12pt]{article}
\newcommand{\eqnum}[1]{\setcounter{equation}{#1}}

\def\be{\begin{equation}}
\def\en{\end{equation}}
\def\bea{\begin{eqnarray}}
\def\ena{\end{eqnarray}}

\newcommand{\square}{\kern1pt\vbox{\hrule height 1.2pt\hbox
{\vrule width1.2pt\hskip 3pt \vbox{\vskip
6pt}\hskip 3pt\vrule width 0.6pt}\hrule height 0.6pt}\kern1pt}

\begin{document}

\renewcommand{\Large}{\large}
\renewcommand{\huge}{\large}

\begin{titlepage}
\baselineskip .15in
\begin{flushright}
WU-AP/32/93
\end{flushright}

{}~\\

\vskip 1.5cm
\begin{center}
{\bf
\vskip 1.5cm
{\large\bf Gauge-invariant Cosmological Perturbations \\[.5em]
in Generalized Einstein Theories}

}\vskip .8in

{\sc Toshinari Hirai}$^{(a)}$ and {\sc Kei-ichi Maeda}$^{(b)}$\\[.5em]
{\em Department of Physics, Waseda University,
Shinjuku-ku, Tokyo 169-50, Japan}
\end{center}
\vfill
\begin{abstract}
Using the covariant approach and conformal transformations,
we present a gauge-invariant formalism for cosmological perturbations in
generalized Einstein theories (GETs), including the Brans-Dicke theory,
theories with a non-minimally coupled scalar field and certain
curvature-squared theories.
We find an enhancement in the growth rate
of density perturbations in the Brans-Dicke
theory, and discuss attractive features of GETs in the structure formation
process.\\
{\em Subject headings}: cosmology: theory --- galaxies: formation ---
relativity --- large-scale structure of universe
\end{abstract}

\vfill
\begin{center}
November, 1993
\end{center}
\vfill
(a)~~electronic mail: 63l512@cfi.waseda.ac.jp\\
(b)~~electronic mail: maeda@jpnwas00.bitnet ~or~ maeda@cfi.waseda.ac.jp\\
\end{titlepage}

\baselineskip .3in

\vspace{.5 cm}
\small
\baselineskip = 15pt


\section{Introduction}\label{sec1}
\par
The formation of structure in the Universe is one of the most interesting
mysteries in nature. The latest observations\cite{COBE} have  revealed that
for galaxy formation, density perturbations must grow much faster than
 the Universe expands. However, the standard theory of
gravitational instability in Einstein's gravity theory does not
provide such a high growth rate. Although some modifications of the
standard theory with appropriate dark matter, a cosmological constant, or
cosmic strings have been proposed,  we do not have any
completely satisfactory solution yet. Hence it is worthwhile to search
for an alternative, in which the gravity sector is modified.

Generalizations of Einstein's theory have found  various interesting
applications to astrophysics\cite{MG6}-\cite{morikawa}. The
Brans-Dicke (BD) theory\cite{BD} or its modified versions may provide a
graceful exit  from the inflationary era of the Universe
(Extended or Hyper-extended inflation\cite{extended}, and soft
inflation\cite{soft}).  Non-minimally coupled terms may produce a
virtual structure in the Universe\cite{morikawa}. There may also exist an
important difference in the growth rate of cosmological density
perturbations, which might resolve the problem of structure formation. In a
preliminary work, Berkin and one of the present authors investigated the
growth of density perturbations in a theory  with a non-minimally coupled
scalar field and showed a sufficient enhancement in the growth of scalar
field perturbations\cite{BerMae}.  Although these authors excluded
ordinary matter fluids and used only a scalar  field, their calculation
suggests that such a coupling may play an important role in  structure
formation in a realistic model with a baryonic matter fluid.

In this paper, we discuss cosmological density perturbations in
generalized  Einstein theories (GETs), which include the
BD theory, theories with a non-minimally coupled scalar field and certain
curvature-squared theories.
The action of GETs we consider here is
 \be
S=\int d^4x \sqrt{-g}\left[ F(\phi, R)-\frac{\epsilon (\phi)}2 (\nabla
\phi)^2
+L_m\right],
\label{Lag}
\en
where $F(\phi, R)$ is an arbitrary function of a scalar field $ \phi $  and
a scalar curvature $R$, ${\epsilon(\phi)} $  is an arbitrary  function of
$ \phi$, and $L_m$ is the Lagrangian of ordinary matter, for which we
assume a perfect fluid.

When we study cosmological perturbations, an appropriate choice of  gauge
becomes important. In order not to pick up unphysical modes by a bad gauge
choice, Bardeen proposed a gauge-invariant (GI) formalism for density
perturbations. Although Bardeen's GI formalism\cite{Bardeen}
is one of the most  elegant and attractive approaches to cosmological
density perturbations, its application to GETs requires tedious
calculations, which have yet to be performed.\footnote {Kodama
and Sasaki nicely reviewed and extended Bardeen's formalism in
\cite{Kodama}. They also tried to formulate GI equations for a theory with
a non-minimally coupled scalar field, but owing
to the complexity of the problem, they made no attempt to solve it
completely.}

On the other hand, Ellis and his collaborators have recently developed
an alternative to  Bardeen's formalism\cite{Ellis1}-\cite{Ellis4}. Their
covariant approach is rather simple. Using their formalism and a conformal
transformation\cite{Whitte}-\cite{Maeda}, we will present GI perturbation
equations in GETs. Although we can write down the basic equations
schematically for an arbitrary GET (see appendix A), we will present
them explicitly only for the following two interesting cases:

{\bf Case (A) : } $F$ is a
linear function of $R$, i.e., $F(\phi, R)=f(\phi)R-V(\phi)$ \hspace{0.5em}
(in \S3).

{\bf Case (B) : } $F$ depends only on  $R$, i.e., $F(\phi, R)=L(R)$
\hspace{0.5em}(in appendix A). \\[.3em]
Case (A) includes the BD theory, induced gravity\cite{induce}, and other
theories with a non-minimally coupled scalar field, while Case (B)
includes
$R^2$-gravity\cite{GETs}.

In order to show the possibility of an enhancement in the growth rate of
density perturbations in GETs, we analyze  perturbations in the BD
theory.

The plan of this paper is as follows:  A brief review of the
covariant approach to density perturbations is given in \S\ref{sec2}. In
\S\ref{sec3} we  derive the perturbation equations for Case (A),
introducing new conformal variables. In \S\ref{sec4} we apply these
equations to BD cosmology and analyze the evolution of
density  perturbations of dust fluid both analytically and numerically.
Finally in
\S\ref{sec5} we discuss general features of density perturbations in GETs.
Appendix A presents the formulation for the most general type of GETs
described by equation (\ref{Lag}) and gives explicit expressions for Case
(B).  In appendix B we show our equations reduce to the already-known ones
in a scalar field dominated universe. In appendix C perturbations in a BD
cosmology are solved analytically.

%
%
%
%
%
\section{Covariant Approach to Cosmological Density Perturbations}
\label{sec2}\eqnum{0}
The covariant approach to cosmological density perturbations,
  introduced originally by Hawking\cite{Hawking} and developed by Ellis
and Bruni\cite{Ellis1}, provides us with a GI formalism, which is an
alternative to Bardeen's approach\cite{Bardeen}-\cite{Kodama}.  Here we
summarize it briefly in order to  define our notation and introduce the
dynamical variables used in this paper.

In the formalism, first we have to select a frame in which GI quantities
are  defined. The frame is described by 4-velocity of an observer, $u^a$.
The time derivative along the motion of the observer (denoted by a dot) of
any rank of tensor ${T^{ab\cdots}}_{cd\cdots}$ and its spatial covariant
derivative in the 3-space $\Sigma$ orthogonal to $u^a$ (denoted by
$^{(3)}{\nabla}_a$) are defined  by
\begin{equation}
{{\dot T}^{ab\cdots}}{}_{cd\cdots}\equiv u^e
\nabla_e{T^{ab\cdots}}_{cd\cdots}, \end{equation}
and
\begin{equation}
^{(3)}{\nabla}_a {T^{bc\cdots}}_{de\cdots}\equiv
{h_a}^m {h_i}^b {h_j}^c\cdots {h_d}^k {h_e}^l\cdots
\nabla_m{T^{ij\cdots}}_{kl\cdots} ,
\end{equation}
respectively, where $ h_{ab}=g_{ab}+u_au_b $ is the
projection tensor into the 3-space $\Sigma$ and
 ${\nabla}_a$ denotes the covariant
derivative with respect to $g_{ab}$.

The derivative of $u^a$ is decomposed as
\begin{equation}
\nabla_b u_a={\omega}_{ab}+{\sigma}_{ab}+\frac 13 \theta h_{ab}-a_a
u_b ,
\label{2.1.4}
\end{equation}
where $a_a \equiv{\dot u}_a$, $\theta \equiv \nabla_a{u^a}$, ${\omega}_{ab}
\equiv ^{(3)}{\nabla}_{[b}u_{a]}$ , and ${\sigma}_{ab}
\equiv ^{(3)}{\nabla}_{(b}u_{a)} $ are the acceleration,
the expansion,
the vorticity tensor and
the shear tensor, respectively.
We define typical length scale $a$ and time scale $H^{-1}$
from the expansion $\theta$ as
\begin{equation}
\frac13 \theta \equiv \frac{\dot a}{a} \equiv  H .
\end{equation}
These $a$ and $H$ turn out to be the scale factor and the Hubble parameter,
respectively, when the Universe is homogeneous and  isotropic. In
inhomogeneous spacetimes, however, those are  local quantities defined by
each observer.

Next we define the GI perturbation variables. The
basic
requirement for GI quantities is that they are invariant under a
general coordinate transformation, i.e., for any choice of
correspondence between a homogeneous and isotropic background
spacetime and the physical, inhomogeneous universe. The simplest GI
quantities are a scalar field which is homogeneous in the background, and
any  vector or tensor field which vanishes in the background. For such
quantities,  all perturbations defined by comparison with
background  values are  free from any gauge dependence.  Such quantities as
$a_a$,
${\omega}_{ab}$ and ${\sigma}_{ab}$ defined in a real lumpy  universe are
GI perturbations, because their background values vanish.

As for the energy density  $\mu$,  the spatial variation
\begin{equation}
X_a\equiv{}^{(3)}{\nabla}_a\mu  \label{315}
\end{equation}
is GI because $\mu$ is homogeneous in the background.
  The GI quantity
\begin{equation}
{\cal D}_a\equiv\frac a{\mu} X_a,
\end{equation}
which is the ratio of its spatial gradient to the density
  at a fixed comoving  scale,
is convenient  in
discussing cosmological perturbations. If we are interested simply in
density perturbations, however, we have to extract the information of its
scalar part from the vector quantity  ${\cal D}_a$, which contains
other information as well.  In this formalism, a local and  unique
splitting\cite{Ellis3} is attained by the operation of the
(comoving) spatial derivative $a^{(3)}{\nabla}_b$ and its decomposition. For
${\cal D}_a$, we have
\begin{eqnarray}
a^{(3)}{\nabla}_b{\cal D}_a &\equiv& {\Delta}_{ab}=\frac13{\Delta}h_
{ab}+{\Sigma}_{ab}+W_{ab} ,\\
{\Delta} \equiv h^{ab}{\Delta}_{ab}
&&W_{ab} \equiv {\Delta}_{[ab]}
,\qquad  {\Sigma}_{ab} \equiv {\Delta}_{(ab)}-\frac13{\Delta}h_{ab} ,\nonumber
\end{eqnarray}
where the skew-symmetric tensor $W_{ab}$ contains information
about vorticity, the symmetric and trace-free tensor ${\Sigma}_{ab}$
describes the evolution of anisotropy  in the universe, and the trace
${\Delta}$ is related  to the aggregation of matter fluid.  This ${\Delta}$,
which corresponds to the density perturbations
$\delta\mu/\mu$, is one of the  most important GI variables\footnote
{${\Delta}$ is related to Bardeen's GI variable
$\epsilon_{m} (= \delta \mu / \mu $ in velocity- orthogonal slicing) in
\cite{Bardeen} as follows\cite{Ellis4}:
$$\Delta=-\sum_n n^2\epsilon_{m}^{(n)}Q^{(n)},\qquad
\epsilon_{m}=\sum_n\epsilon_{m}^{(n)}Q^{(n)},$$ to first order of
perturbations, where $n$ is a wave number of the harmonics
$Q^{(n)}$.}.

Obtaining the evolution equations for GI variables is rather simple.
First, take spatial derivatives  (by operating $a^{(3)}{\nabla}_a$) of
 the fundamental equations, which consist of the energy conservation, the
momentum conservation, the  Raychaudhuri, and the Gauss-Codacci
equations (see e.g.\cite{RC} or \cite{cargese}). Then, take the (comoving)
divergence  (by operating $a^{(3)}{\nabla}^a$ to the gradient equations) and
construct scalar equations for GI variables such as $\Delta$ defined
above. The  resultant equations describe the evolution of GI  scalar
perturbations.

%
%
%
%
%
\section{Formulation of Perturbation Equations in GETs}\label{sec3}\eqnum{0}
\subsection{Conformal Transformation}
\par
We now derive GI perturbation equations in GETs, using the covariant
formalism described in \S\ref{sec2}. In this section, we consider Case (A),
i.e., the case where $F(\phi, R)$ in equation (\ref{Lag}) is a linear
function of
$R$.  As for Case (B), i.e., the case where $F(\phi, R)$ depends only on
$R$, we will present explicit perturbation equations in appendix A, which
turn out to be the same as those given here.
Furthermore, the derivation of perturbation equations for  the most
general action (\ref{Lag}) is straightforward (see also appendix A).

The theories we consider in this section are described by the
action
\begin{equation}
S=\int d^4x \sqrt{-g}\left[ f(\phi)R-\frac{\epsilon (\phi)}2 (\nabla
\phi)^2-V(\phi)+L_m\right],
\label{3.2.1}
\end{equation}
where $f(\phi)$, ${\epsilon(\phi)} $ and $V(\phi)$ are arbitrary
functions of $\phi $.  $L_m$ is the Lagrangian of ordinary matter, for
which we  assume a perfect fluid, i.e.,
\begin{equation}
T_{ab}^{M}={\mu} u_a^{M} u_b^{M}+p h_{ab}^{M}
\label{perfect},
\end{equation}
where $\mu$, $p$ and $u^M_a$ represent the energy density, the pressure
and the 4-velocity of the matter fluid, respectively, and $h^M_{ab}$ is the
projection tensor into the 3-space orthogonal to $u^M_a$. Taking variations
of equation (\ref{3.2.1}) with respect to the metric
$g_{ab}$ and
$\phi$, we find  the basic equations
 \bea
G_{ab}&=&\frac1{f(\phi)}\left\{\frac{\epsilon(\phi)}2\left[{\nabla}_a\phi
{\nabla}_b\phi-\frac12g_{ab}({\nabla}\phi)^2 \right]+\left[ {\nabla}_a
{\nabla}_b f(\phi)-\square f(\phi)g_{ab}\right] \right. \nonumber \\
&&-\left. \frac12 V(\phi)g_
{ab}+
\frac12 T_{ab}^{M}\right\},
\label{3.2.2}
\ena
\bea
&&\hspace{-1.5em}\epsilon(\phi)\square\phi+\frac12
\frac{d\epsilon(\phi)}{d\phi}(\nabla\phi)^2 - \frac{dV(\phi)}{ d
\phi} \nonumber \\
&&\hspace{2.5em}+\frac{df(\phi)/d\phi} {f(\phi)}
\left\{3\square f(\phi)+\frac{\epsilon(\phi)}2({\nabla}\phi)^2 +2V(\phi)+
\frac12 ({\mu}-3p) \right\}=0, \label{3.2.3}
\ena
where $G_{ab}= R_{ab} - \frac{1}{2} g_{ab}R$ is the Einstein tensor.

The application of the covariant approach in \S 2 to the present model
is not straightforward because of
 the presence of
higher-derivative  terms such as ${\nabla}_a{\nabla}_b f(\phi)$ and $\square
f(\phi)g_ {ab}$.
Although there may be a way to resolve this problem leaving the present
variables as they are, we prefer to circumvent it by changing variables
using the conformal transformation\cite{Maeda}
\begin{equation}
{\hat{g}}_{ab}=e^{2\omega(x)}g_{ab}. \label{3.3.0}
\end{equation}
 If we choose the conformal factor as
\begin{equation}
e^{2\omega}=2{\kappa}^2|f(\phi)| ,
\end{equation}
the field equations become
\begin{eqnarray}
{\hat{G}}_{ab} &=& {\kappa}^2 \left\{{\hat{\nabla}}_a\varphi {\hat
{\nabla}}_b\varphi-\frac12{\hat{g}}_{ab}(\hat{\nabla}\varphi)^2 -U
(\varphi){\hat{g}}_{ab} \right\}+{\kappa}^2N(\varphi) \left\{{\mu}
\hat{u}_a^{M} \hat{u}_b^{M}+p \hat{h}_{ab}^
{M}\right\} \nonumber \\
& \equiv & {\kappa}^2(\hat{T}_{ab}^{\varphi}+\hat{T}_{ab}^{M})
,\label{3.3.1}\\
\stackrel{\hat{}}{\square
} \varphi &-& U'(\varphi)+
{\kappa}^2 M(\varphi)\{{\mu}-3 p\}=0 ,
\label{3.3.2}
\end{eqnarray}
with
\begin{eqnarray}
{\kappa}\varphi &=& \int d\phi \left[\frac{{\epsilon}(\phi)f(\phi)+3(df
(\phi)/d\phi)^2}{2f^2(\phi)}\right]^{\frac12} ,\\
U(\varphi) &=& \frac{(\mbox{sign})}{(2{\kappa}^2 f(\phi))^2}V(\phi)
 ,\qquad \qquad (\mbox{sign})=\frac{f(\phi)}{|f(\phi)|} ,\\
M(\varphi) &=& \frac{df(\phi)/d\phi}{(2{\kappa}^2 f(\phi))^2 [2{\kappa}^2
\{ {\epsilon} (\phi) f(\phi)+3(df(\phi)/d\phi)^2\}]^{\frac12}} ,\\
N(\varphi) &=& \frac{(\mbox{sign})}{(2{\kappa}^2 f(\phi))^2} ,
\ena
\bea
{\hat{u}}_a^{M} &=& e^{\omega}u_a^{M} ,\qquad
{\hat{u}}^a_{M} = e^{-\omega}u^a_{M} ,\\
{\hat{h}}_{a}^{Mb} &=& h_{a}^{Mb }\label{3.3.3} ,\qquad
{\hat{h}}_{ab}^{M} = e^{2\omega} h_{ab} ^{M} ,\qquad
{\hat{h}}^{ab}_{M} = e^{-2\omega} h^{ab}_{M} ,
\end{eqnarray}
where $\kappa^2=8\pi G$, $'$~(a prime)~=$d/d\varphi$, and new variables
with
{}~$\hat{}$~(a caret) denote those with respect to ${\hat{g}}_ {ab}$. The
resultant basic equations in terms of new variables turn out to be
familiar ones, that is, those for the Einstein gravity $(\hat{g}_{ab})$
plus   a minimal scalar field $\varphi$ with a potential $U(\varphi)$ and
a perfect fluid, which interact with each other through coupling functions
$M(\varphi)$ and
$N(\varphi)$.  The non-minimal coupling to a scalar curvature has been absorbed
into such interactions. Note that we have not changed the
original energy density $\mu$ and pressure $p$ for a perfect fluid.

We can recover  physical  variables from new ones through equation
(\ref{3.3.0}), which gives the relations between
the length scales and  Hubble parameters

\bea
\hat{a}&=&e^{\omega}a \label{scale} ,\\
\hat{H}&=&e^{-\omega}(H+\dot{\omega}) \label{Hubble} ,
\ena
and equations (\ref{3.3.3}). In particular, the density perturbation
$\Delta^{M}$ observed along fluid flow $u_a^{M}$ is obtained from
$\hat{\Delta} ^{M}$ as follows:
First, we have the relation
\begin{equation}
{\cal{D}}_a^{M} \equiv \frac a{\mu}{}^{(3)}{\nabla}_a^{M}\mu=
\frac a{\mu}h_{a}^{M b}{\nabla}_b \mu=
e^{-\omega}\frac {\hat{a}}{\mu}{\hat{h}_{a}}^{M b}{\hat{\nabla}}_b \mu=
e^{-\omega}\frac {\hat{a}}{\mu}{}^{(3)}{\hat{\nabla}}_a^{M}\mu \equiv
e^{-\omega}\hat{\cal{D}}_a^{M},  \label{trans}
\end{equation}
to first order in the perturbations.
Taking the comoving divergence of this relation by $a{^{(3)}}{\nabla}^a_
{M}(=e^{\omega}\hat{a}{}^{(3)}{\hat{\nabla}}^a_{M})$ yields
\begin{equation}
\Delta^{M}=\hat{\Delta} ^{M}, \label {scaG}
\end{equation}
Thus, there is an advantage to using the
conformal variables. The original covariant approach is applied as it is
to the simplified conformal equations and it gives the same density
perturbations as those in terms of the physical variables. In what follows,
we will formulate equations in terms of the new conformal variables. We
omit  ~$\hat{}$~(a caret) for brevity.  When we discuss the results, of
course, we transform back to the physical variables (see \S
4).

\subsection{Formulation in terms of  Conformal Variables }\label{sec3.2}
\par
The total energy-momentum tensor, $T^{*}_{ab}\equiv T_{ab}^{\varphi}+T_
{ab}^{M}$ in equation (\ref{3.3.2}), takes on a fluid form for any
observer with 4-velocity
$u_a^{O}$\cite{Ellis5},
\be
T^*_{ab}=\mu_* u_a^{O} u_b^{O} + p_* h^{O}_{ab}+q^{O}_{*(a } u_{ b)}^{O} +
{\pi}^{O}_{*ab} ,
\label{3.4.1}
\en
with
\bea
\mu_* & \equiv & \frac12 {\dot{\varphi}}^2+U(\varphi)+N(\varphi){\mu} ,
\label{3.4.2} \\
p_* & \equiv & \frac12 {\dot{\varphi}}^2-U(\varphi)+N(\varphi)p ,\label
{3.4.3}
\\
q^{O}_{*a} & \equiv &
-{\dot{\varphi}}^{(3)}{\nabla}_a\varphi+N(\varphi)({\mu}+ p)V_a^{M} ,\\
{\pi}^{O}_{*ab} &=& 0 ,
\end{eqnarray}
where we have neglected the higher than first order of
perturbations composed of all the spatial derivatives and the  relative
velocity of the matter fluid with respect to
$u_a^{O}$,
\begin{equation}
V_a^{M} \equiv u_a^{M}-u_a^{O}.
\end{equation}
In what follows, we will also neglect higher-order perturbations in all
equations, as in the usual first order perturbation theory.
In equation (\ref{3.4.1}), $\mu_* $,
$p_* $, $q^{O}_{*a} $ and ${\pi}^{O}_{*ab} $ represent the total energy
density,  pressure, energy flux and anisotropic pressure measured by the
observer $u_a^{O}$. Note that the energy density $\mu_*
$ and the pressure $p_* $ do not depend on the observer, but only the
energy flux does.

There is  no physical constraint on $u_a^{O}$ other than that $V_a^{M}$ be
small. However, there exists  a  preferred choice, i.e.,
the  4-velocity of the center of mass. In this reference frame,
\ the total energy
flux vanishes. Ellis et al. call it the energy frame
and we use $u_a^E$ to denote its 4-velocity.
The total energy-momentum tensor, written in terms of the center of mass
observer ($u_a^E$), takes the perfect fluid form:
\begin{equation}
T_{ab}=\mu_* u_a^E u_b^E +p_*h_{ab}^E, \label{total}
\end{equation}
where $\mu_*$ and $p_*$ are defined by equations (\ref{3.4.2}) and
(\ref{3.4.3}).

Once a reference frame is fixed,  we can derive  explicitly the evolution
equations from
energy conservation, momentum conservation, the Raychaudhuri equation, the
equation of motion for $\varphi$, and the Gauss-Codacci equation in the
center of mass frame, which are
\begin{equation}
\dot{\mu}_* + 3H(\mu_* +p_*) =0 ,
\label{3.5.1}
\end{equation}
\begin{equation}
(\mu_* +p_*) a_a+^{(3)}{\nabla}^E_ap_*=0 ,
\label{3.5.2}
\end{equation}
\begin{equation}
3\dot {H}+3H^2 -{\nabla}_a{a^a}+\frac 12 {\kappa}^2(\mu _*+
3p_*) +2(\sigma^2-\omega^2)=0 ,
\label{3.5.3}
\end{equation}
\begin{equation}
^{(3)}R =2(-\frac13 {\theta}^2+\mu_*+\sigma^2-\omega^2),
\label{3.5.5}
\end{equation}
\begin{equation}
 \square \varphi -U'(\varphi)+{\kappa}^2M (\varphi)({\mu} -3p) =0 ,
\label{3.5.4}
\end{equation}
where $\sigma^2\equiv \frac12 \sigma^{ab}\sigma_{ab}$ and
$\omega^2\equiv \frac12 \omega^{ab}\omega_{ab}$.

The equations for the background isotropic and homogeneous fields
are derived by
setting $\sigma$, $\omega$ and the spatial derivatives equal to zero
in equations (\ref{3.5.1})$\sim$(\ref{3.5.5}) as
\begin{equation}
\dot{\mu}+3H({\mu}+p)+\frac{N'(\varphi)}{N(\varphi)}\mu\dot{\varphi}
+{\kappa}^2\frac{M(\varphi)}{N(\varphi)}({\mu}-3p)\dot{\varphi}=0
,\label{3.31}
\end{equation}
\begin{equation}
3\dot {H}+3H^2+{\kappa}^2(\dot{\varphi}^2-U(\varphi)) +\frac 12
{\kappa}^2N(\varphi)(\mu +3p) =0 ,
\end{equation}
\begin{equation}
3H^2+\frac{3k}{a^2}={\kappa}^2\left(\frac12 {\dot{\varphi}}^2+U
(\varphi)+N(\varphi){\mu}\right) ,
\end{equation}
\begin{equation}
\ddot{\varphi}+3H\dot{\varphi} +U'(\varphi)-
{\kappa}^2M(\varphi)({\mu} -3p) =0 ,\label{3.34}
\end{equation}
where $k (=0$ or $\pm 1)$ is  the so-called
spatial curvature constant in the
Friedmann-Lema\^{\i}tre-Robertson-Walker(FLRW) universe.

The scalar GI perturbation variables for the matter and scalar field in the
center of  mass frame are defined as
\begin{eqnarray}
\Delta_*^E  &=& \frac{a^2}{{\mu}_*}{}^{(3)}{\nabla}_E^2{\mu}_*,\qquad
{\cal{Y}}_*^E =  a^2{}^{(3)}{\nabla}_E^2{p}_*,\\
{\Delta}^E &=&\frac{a^2}{{\mu}}{}^{(3)}{\nabla}_E^2{\mu} ,\qquad
{\cal{Y}}^E={a^2}^{(3)}{\nabla}_E^2{p} ,\\
{\cal{Z}}^E &=& a^2{}^{(3)}{\nabla}_E^2{\theta}\\
{\Phi}^E &=& {a^2}^{(3)}{\nabla}_E^2{\varphi}.
\end{eqnarray}

Following the procedure in \S 2, we find the first order perturbation
equations as follows:
\bea
\dot{\Delta}_*^E & - & H\frac {p_*}{{\mu}_*}{\Delta}_*^E+(1+\frac{ p_*}
{{\mu}_*}){\cal Z}^E=0 ,
\label{442a}\\
\dot{\cal Z}^E & + & 2H{\cal Z}^E+\frac12{\kappa}^2 \mu_*{\Delta}_*^E+
\frac 1{\mu_*+p_*} \left( \frac k{a^2}+{}^{(3)}{\nabla}^2_E\right){\cal
Y}_*^E=0 .
\label{442b}\\
\ddot{\Phi}^E & + & 3H\dot{\Phi}^E+\left\{ U''(\varphi) -{\kappa}^2
\left(M'(\varphi) \right) (\mu-3p) \right\}
{\Phi}^E \nonumber \\
&&\quad -{\kappa}^2  M(\varphi)\mu \Delta^E+3
{\kappa}^2 M(\varphi){\cal{Y}}^E+\dot{\varphi}{\cal{Z}}^E-
{}^{(3)}{\nabla}^2_E{\Phi}^E-a^2\dot{\varphi}{}^{(3)}{\nabla}^a_E\dot
{a_a} \nonumber \\
&&\quad -a^2(4H\dot{\varphi}+2\ddot{\varphi}){}^{(3)}{\nabla}^a_E
a_a=0\label{442c}.
\ena
Here we have used the following useful relations for
$\omega_{ab}$, $\varphi$, an arbitrary vector $\Upsilon_a^E$ orthogonal
to  $u^a_E$ ($\Upsilon_a^E u^a_E=0$) and
$\Upsilon^E \equiv a{}^{(3)}{\nabla}^a_E \Upsilon_a^E$:
\bea
&&a{}^{(3)}{\nabla}^a_E{}^{(3)}{\nabla}^b_E \omega_{ab}=0 , \\
&&\square\varphi=\frac1{a^2}{\Phi}^E-3H\dot{\varphi}-\ddot{\varphi},\\
&&a{}^{(3)}{\nabla}^a_E \dot{\Upsilon}_a^E=\dot{\Upsilon}^E,
\quad\qquad a{}^{(3)}{\nabla}^a_E
\ddot{\Upsilon}_a^E=\ddot{\Upsilon}^E,\nonumber \\
&&a{}^{(3)}{\nabla}^a_E\left[({}^{(3)}{\nabla}^2_E-\frac{2k}{a^2})
\Upsilon_a^E\right]={}^{(3)}{\nabla}^2_E\Upsilon^E ,
\ena
to first order of perturbations. See the appendix of \cite{Ellis3} for
more details.

The explicit equations in terms of ${\Delta}^E$, ${\cal{Y}}^E$, ${\Phi}^E$ and
${\cal{Z}}^ E$ are derived from equations (\ref
{442a})$\sim$(\ref{442c}), by
using relations (\ref{3.4.2}), (\ref{3.4.3}) and
\begin{eqnarray}
\Delta_*^E &=& \frac1{\frac12 {\dot{\varphi}}^2+U(\varphi)+N(\varphi)
{\mu}}\left[N(\varphi){\mu}{\Delta}^E-a^2\dot{\varphi} {}^{(3)}
{\nabla}^a_E a_a
\right.  \nonumber \\
 & & \left.+\dot{\varphi}\dot{\Phi}^E+\left(U'(\varphi)
+{\mu}N'(\varphi) \right){\Phi}^E \right].
\label{3.6.1}
\end{eqnarray}
Equation (\ref{3.5.2}) is also used to eliminate the acceleration $a_a$.
To describe the resulting equations, we expand the
dynamical variables as
\begin{equation}
\Delta=\sum_n {\Delta}_{(n)}(t) Q^{(n)},
\end{equation}
where $Q^{(n)}$ is an eigenfunction of the Laplace-Beltrami
operator ${}^{(3)}{\nabla}^2$\cite {Ellis4}:
\begin{equation}
^{(3)}{\nabla}^2Q^{(n)}=-\frac{n^2}{a^2}Q^{(n)}.\label{harmo}
\end{equation}

Finally, we obtain the basic equations for the harmonic
components of perturbed quantities as follows (We drop the suffix
$(n)$ for brevity):
\begin{eqnarray}
\dot{\Delta}^E=
&&3w\left(H-\kappa^2\frac{M}{N}\dot{\varphi}\right) \Delta^E
+3{\kappa}^2\frac{M}{N}\frac{\dot{\varphi}}{\mu}{\cal{Y}}^E
-\left(1+w\right){\cal{Z}}^E \nonumber \\
&+&\left[\frac{3H\dot{\varphi}}{N\mu}-
{\kappa}^2 \frac{M}{N}(1-3w)-\frac{N'}{N} \right]\dot{\Phi}^E \nonumber \\
&+&\left[-\frac{3H}{N\mu}\left(U'-N'\mu w\right)
+\dot{\varphi}\left\{\frac{n^2}{a^2}\frac{1}{N\mu}
-\left(\frac{N'}{N}\right)^{'}
-{\kappa}^2\left(\frac{M}{N}\right)^{'}(1-3w)\right\}\right]
\Phi^E  ,\label{delta} \\ \nonumber \\
\dot{\cal{Z}}^E=
&-&\frac12{\kappa}^2N{\mu}{\Delta}^E
-\frac{1}{2\mu(1+w)}\left({\kappa}^2{\dot{\varphi}}^2-\frac{2n^2}
{a^2}+\frac{6k}{a^2}\right){\cal{Y}}^E
- 2H{\cal{Z}}^E\nonumber \\
&-&\frac{\dot{\varphi}}{2}
\left[{\kappa}^2{}+\frac{1}{N\mu(1+w)}\left({\kappa}^2{\dot
{\varphi}}^2-\frac{2n^2}{a^2}+\frac{6k}{a^2}\right)\right]\dot{\Phi}^E
\nonumber \\
&-&\frac12\left[{\kappa}^2(U'+N'{\mu})-\frac{(U'-N'\mu
w)}{N\mu(1+w)}\left({\kappa}^2{\dot{\varphi}}^2-\frac{2n^2}
{a^2}+\frac{6k}{a^ 2}\right)\right]\Phi ^E, \label{z}\\ \nonumber \\
\ddot{\Phi}^E=
&-&\frac{1}{\dot{\varphi}^2+N\mu(1+w)}
\Biggl[3(H+{\kappa}^2\frac{M}{N}\dot{\varphi})N\mu(1+w)
+\frac{1+c_s^2}{1+w}{\dot{\varphi}}^3
\left\{\frac{N'}{N}+\kappa^2\frac{M}{N}(1-3w)\right\}\Biggr.\nonumber
\\ & &\Biggl. -\frac{N'}{N}{\dot{\varphi}}^3
+\dot{\varphi}\left\{3(c_s^2-1)H{\dot{\varphi}}
-4U'-12{\kappa}^2M\mu w +N'\mu w\right\}\Biggr]\dot{\Phi}^E\nonumber\\
&-&\frac{1}{\dot{\varphi}^2+N\mu(1+w)}
\Biggl[\left\{U''-{\kappa}^2M'
\mu(1-3w)+\frac{n^2}{a^2}+\frac{3c_s^2HN'\dot{\varphi}}{N}\right\}N\mu(1+w)
\Biggr.\nonumber \\
& &-\Biggl.(U''-N''\mu w){\dot{\varphi}^2}
+c_s^2\frac{N'}{N}\mu\dot{\varphi}^2\left\{N'+\kappa^2M(1-3w)\right\}
\Biggr.\nonumber \\
&&+\Biggl.(U'-N'\mu w)
\Biggl\{\frac{N'}{N}\frac{w-c_s^2}{1+w}\dot{\varphi}^2
-\kappa^2\frac{M}{N}(1-3w)\left(2N\mu+\frac{1+c_s^2}{1+w}\dot{\varphi}^2
\right)\Biggr.\Biggr.\nonumber\\
&&-\Biggl.\Biggl.(3c_s^2H{\dot{\varphi}}-2U')
\Biggr\}\Biggr]\Phi^E\nonumber\\
&+&\frac{{\kappa}^2MN\mu^2(1+w)}{\dot{\varphi}^2+N\mu(1+w)}
{\Delta}^E
-\frac{N\dot{\varphi}}{\dot{\varphi}^2+N\mu(1+w)}\dot{\cal{Y}}^E\nonumber\\
&-&\frac{1}{\dot{\varphi}^2+N\mu(1+w)}
\Biggl[ {\kappa}^
2MN\mu(5-3w)+N(3c_s^2H{\dot{\varphi}}-2U')
\Biggr.\nonumber \\
&&+\Biggl.\frac{1+c_s^2}{1+w}\dot{\varphi}^2
\left\{N'+\kappa^2M(1-3w)\right\}\Biggr] {\cal{Y}}^E \nonumber \\
&-&\frac{N\mu(1+w)\dot{\varphi}}{\dot{\varphi}^2+N\mu(1+w)}
{\cal Z}^E ,\label{phi}
\end{eqnarray}
where we have used notations $w\equiv p/\mu$, $c_s^2 \equiv
\dot{p}/\dot{\mu}$\cite{Bardeen}.

These complicated equations may be much simplified in the case where
$N(\varphi)$, $M(\varphi)$ and/or $U(\varphi)$ are
expressed in terms of exponential functions of $\varphi$. This happens in
several interesting models (see eqs.
(\ref{reddelta})$\sim$(\ref{redphi}) for the BD theory and also appendix
A).

Although the variables in the center of mass frame are easy to handle, when
we discuss the density perturbations of matter fluid, we should
 evaluate the density fluctuations of the matter fluid in the matter frame
defined by $u_a^{M}$. Hence, we must construct the density perturbations
in the matter frame from variables in the center of mass
frame\cite{Ellis5}. First, the  projection tensor
$h_{ab}^{M}$  defined by $u_a^{M}$ and the energy density of matter
${\mu}^{M}$  in the matter frame are written as
\begin{eqnarray}
h_{ab}^{M}&=&g_{ab}+u_a^{M}u_b^{M} =h_{ab}+u_a^EV_b^{M}+
V_a^{M}u_b^E ,\nonumber \\
{\mu}^{M}&=&{\mu} \nonumber,
\end{eqnarray}
to first order. Then, using these relations and the condition
$$
V_a^{M}=\frac{\dot{\varphi}}{aN({\mu}+ p)}{\Phi}^E_a,
$$
which is derived from the condition  that the total energy
flux vanishes in the  center of  mass frame,  the density perturbation
 in the matter frame
${\Delta}^{M}$ is  expressed as
\begin{eqnarray}
{\Delta}^{M}& \equiv &{a}{}^{(3)}{\nabla}_{M}^a({\cal{D}}_a^{M})=ah_
{M}^{ab}{\nabla}_b(\frac a{{\mu}^{M}}h_{a}^{Mc}{{\nabla}_c}
{\mu}^{M}) \nonumber \\
&=&{\Delta}^E+\frac{\dot{\mu}}{{\mu}}\frac{\dot{\varphi}}{N({\mu}+
p)}\Phi^E .
\end{eqnarray}

In some specific theories, we recover already-known basic
equations. For example, setting
$f(\phi)=1/2\kappa^2$, and $\epsilon(\phi)=V(\phi)=0$, we find
immediately the perturbation equations for the dust universe in the
Einstein theory. Hwang\cite{Hwang} and Bruni et al.\cite{scalar} derived
perturbation equations in some scalar field dominated universes. We can also
recover these equations by taking the limit $\mu$, $p\rightarrow 0$  and
rewriting our variables in terms of their variables (see appendix B).

%
%
%
%
\section{Density Perturbations in the Brans-Dicke Theory}
\label{sec4}\eqnum{0}
 In order to see how the coupling term to the curvature  scalar  in GETs
changes the growth rate of density perturbations, we consider the BD
theory\cite{BD} as an example. In the BD theory, the reciprocal of the
scalar field,
$1/\phi$, corresponds to an ``effective" gravitational constant $G(\phi)$.
If a flat dust universe ($k=0$, $p=0$) satisfies a specific initial
condition
\be
\frac{\mu_0t_0^2}{\phi_0}=\frac{(3+2\omega)}{4\pi(4+3\omega)},\label{special}
\en
where $\omega$ is a constant (the BD parameter) and the suffix $0$
represents the present time, the BD scalar field
$\phi$, the energy density of matter $\mu$ and the scale $a$ evolve in
time as
\bea
\phi &=& \phi_0\left(\frac t{t_0}\right)^{2/(4+3\omega)}, \label{spephi}\\
a &=& \left(\frac t{t_0}\right)^{(2+2\omega)/(4+3\omega)},\\
\mu &=& \mu_0a^{-3}=\mu_0
\left(\frac t{t_0}\right)^{-3(2+2\omega)/(4+3\omega)}.\label{spemu}
\ena
In the limit of large $\omega$, this solution approaches
the Einstein de-Sitter universe ($\phi=\mbox{constant}$, $a\propto
t^{2/3}$). When the condition (\ref{special}) is not satisfied
initially, the behaviors of
$\phi$ and $a$ in the early stage may differ considerably from the above
solution for the same value of $\omega$\cite{Weinberg}.

 Nariai\cite{Nariai} analyzed density perturbations in
the above specific solution (\ref{spephi})$\sim$(\ref{spemu}) of the BD
cosmology using a specific gauge condition and solved its evolution
equations analytically as
\bea
{\delta\mu \over \mu} & = & \frac {c_1}{(1+\nu)(1+2\nu)}
\left\{ \frac12(2\nu nt_0)^2 \left(\frac t{t_0}\right)^{\frac1{\nu}}
+(3-2\nu)(1+\nu)\right\}+c_2 \left(\frac t{t_0}\right)^{-1}\nonumber\\
&&+\left(\frac
t{t_0}\right)^{-\frac12}\{c_3J_{\nu}(nT)+c_4J_{-\nu}(nT)\}\label{dependent},
\ena
where $c_1$, $c_2$, $c_3$ and
$c_4$ are arbitrary integration constants, $\nu \equiv
(4+3\omega)/2(2+\omega)$,
$T\equiv2\nu t_0 ( t/{t_0})^{-\frac1{2\nu}}$, $J_{\nu}$ is the Bessel
function of the $\nu$-th order and $n$ represents the wave number of
harmonics.

In our formalism, the BD theory is given by setting
\bea
f(\phi)&=&\frac{\phi}{16\pi},\\
\epsilon(\phi)&=&\frac{\omega}{8\pi \phi},\\
V(\phi)&=&0,
\ena
in the action (\ref{3.2.1}). New variables, potential and coupling
functions in the conformal frame introduced in \S3 are given as
\begin{eqnarray}
\frac{d\hat{t}}{dt}&=&\sqrt{\frac{\phi}{\phi_0}},\\
\hat {a} &=&\sqrt{\frac{\phi}{\phi_0}}a ,\\
\varphi &=&\frac{\sqrt{2(3+2\omega)}}{2\kappa}
\ln\left|\frac{\phi}{\phi_0}\right|,\\
N(\varphi)&=&\exp\left[-\frac{4\kappa}{\sqrt{2(3+2\omega)}}\varphi\right],\\
M(\varphi)&=&\frac1{\kappa\sqrt{2(3+2\omega)}}N(\varphi),\\
U(\varphi) &=&0.
\end{eqnarray}
The exponential form of $N$ and $M$ and vanishing $U$  rather simplify the
perturbation equations (\ref{delta})$\sim$(\ref{phi}). Moreover, if we
assume the ordinary matter is dust ($p=0$), they are reduced as
\begin{eqnarray}
\dot{\Delta}^E=&-&{\cal{Z}}^E
+\left[\frac{3H\dot{\varphi}}{N\mu}+
\frac{3\kappa}{\sqrt{2(3+2\omega)}}\right]\dot{\Phi}^E \nonumber \\
&+&\frac{n^2 \dot{\varphi}}{a^2N\mu}{\Phi}^E,\label{reddelta}\\
\dot{\cal{Z}}^E= &-&\frac12{\kappa}^2N\mu
{\Delta}^E-2H{\cal{Z}}^E\nonumber \\
&-&\frac{\dot{\varphi}}{2}\left[{\kappa}^2 +\frac1{N\mu}
\left({\kappa}^2{\dot{\varphi}}^2-\frac{2n^2} {a^2}+\frac{6k}{a^2}\right)
\right]\dot{\Phi}^E \nonumber \\
&+&\frac{2\kappa^3N\mu}{\sqrt{2(3+2\omega)}}
 \Phi ^E,\\
\ddot{\Phi}^E= &-&\frac{1}{\dot{\varphi}^2+
N\mu}
\left[3H(N\mu-{\dot{\varphi}}^2)
+\frac{\kappa\dot{\varphi}}{\sqrt{2(3+2\omega)}}
(3N\mu+{\dot{\varphi}}^2)\right]\dot{\Phi}^E\nonumber\\
&-&\frac{N\mu } {\dot{\varphi}^2+N\mu }
\left[\frac{2{\kappa}^2 N\mu }{3+2\omega}
+\frac{n^2}{a^2}\right]\Phi^E\nonumber\\
&+&\frac{\kappa N^2\mu^2}
{\dot{\varphi}^2+N\mu}
\frac1{\sqrt{2(3+2\omega)}}{\Delta}^E
-\frac{N\mu\dot{\varphi}}
{\dot{\varphi}^2+N\mu}{\cal Z}^E .\label{redphi}
\end{eqnarray}
In the case of a flat dust universe ($k=0$, $p=0$) with the specific
initial condition (\ref{special}), we can solve these equations analytically
(see appendix C). As for the density perturbation of dust, we obtain
\bea {\Delta}^E=&&
c_1\left\{\left(\frac{\hat{t}}{t_0}\right)^{\frac{2(2+\omega)}{5+3\omega}}
+\frac{4(8+5\omega)}{(5+3\omega)^2}
\left(\frac{5+3\omega}{4+3\omega}\right)^{\frac{1+\omega}{5+3\omega}}
\frac1{{t_0}^2 n^2}\right\}\nonumber \\
&+&c_2\left(\frac{5+3\omega}{4+3\omega}\frac{\hat{t}}{t_0}\right)
^{-\frac{4+3\omega}{5+3\omega}}\nonumber \\
&+&c_3 \left(\frac{\hat{t}}{t_0}\right)^{-\frac{4+3\omega}{2(5+3\omega)}}
J_{\nu}(nT) +c_4 \left(\frac{\hat{t}}{t_0}\right)^{-\frac{4+3\omega}
{2(5+3\omega)}}
J_{-\nu}(nT),\label{soldelta}
\ena
where $c_1$, $c_2$, $c_3$ and $c_4$ are arbitrary integration constants, and
\be
\nu\equiv \frac{4+3\omega}{2(2+\omega)},\qquad
T\equiv\frac{4+3\omega}{2+\omega}
\left(\frac{5+3\omega}{4+3\omega}\frac{\hat{t}}{t_0}\right)
^{\frac{2+\omega}{5+3\omega}},
\en
which are the same as Nariai's notation. We can easily show that
this solution is essentially the same as
 the gauge-dependent one (\ref{dependent}) by use of the
relation (\ref{tthat}).

The density perturbation $\Delta^E$ has four independent modes. The first
two terms have their counterparts in the Einstein de-Sitter model in the
limit of $\omega \rightarrow \infty$, while the last two originate from the
BD scalar field. The growing mode (the first term in equation
(\ref{soldelta})) evolves asymptotically
\be
\left(\frac{\hat{t}}{t_0}\right)^{\frac{2(2+\omega)}{5+3\omega}}
\propto \left(\frac{t}{t_0}\right)^{\frac{2(2+\omega)}{4+3\omega}}
\propto a^{\frac{2+\omega}{1+\omega}} ,
\en
which shows that the growth rate of $\Delta^E$ in this
specific BD model is somewhat higher (by $a^{\frac1{1+\omega}}$ times) than
that in the Einstein de-Sitter model.

Although we have obtained an analytic solution for the above specific model,
in order to see whether such an enhancement in the growth rate of density
perturbations is generic and to find what the  key to
such an enhancement is, we reanalyze numerically the density perturbations
 of BD cosmological models with arbitrary initial conditions using
our formalism.

We also evaluate the growth of density perturbations quantitatively
under constraints from observations. Thus we can confirm that if  all
 the observations are reliable, then the enhancement in the BD theory
itself is inadequate to resolve the structure formation problem. The
observational constraints on the BD theory may be summarized as follows:\\
\noindent  {\bf1}) The BD parameter\cite{viking};
$$\omega > 500,$$

\noindent {\bf2}) The present rate  of time variation of $G$\cite{pulsar};
$$ \left| \frac {\dot{G}}{G}\right| _{present}\le 1\times 10^{- 11}\mbox
{year}^{-1},$$

\noindent {\bf3}) From nucleosynthesis\cite{Accetta};
$$\left| \frac {\Delta G}{G}\right| _{nucleosynthesis}\le 40\%,$$

where $\Delta G \equiv  G - G|_{present}$.

Among these constraints, {\bf1}) and {\bf2}) are obtained directly from
present-day observations of gravity, while {\bf3}) is a result obtained
indirectly, i.e., through the comparison between the theoretically
predicted and observationally inferred primordial abundances of light
elements. If we adopt {\bf3}), which critically depends on the theory of
nucleosynthesis, the time variation of  gravitational constant  in the past
is also strongly restricted, resulting in no difference in the evolution
of  density perturbations from that in the Einstein theory, as we will see
shortly.

For the sake of easy comparison with the Einstein de-Sitter model,
we first assume the ordinary matter is dust ($p=0$) and the present
density parameter $\Omega_{dust,0}=\kappa^2\mu_0/3H_0^2$ equals unity,
which means the universe is closed ($k>0$) in the BD cosmology. Besides,
we set the Hubble constant $H_0 =100\mbox{km/sec/Mpc}$. Then, regarding the
observational constraints, we consider the following two cases:

{\bf Case (i) : } All the constraints {\bf1}), {\bf2}) and {\bf3}) are
satisfied.

{\bf Case (ii) : } Only {\bf1}) and {\bf2}) are satisfied.

\noindent
In all the calculations below, we set the BD parameter $\omega=500$,
therefore the constraint {\bf1}) is always satisfied. As for constraint
{\bf3}), we  calculated the background spacetime back to the
nucleosynthesis era where $a\sim 10^{-9}$ and found
$|\dot{G}/G|_{present} \le 2.039\times 10^{-13}\mbox{year}^{-1} $  for the
universe which satisfies the constraint
$|{\Delta G}/{G}| _{nucleosynthesis}\le 40\%$. Hence we adopt
$|\dot{G}/G|_{present} =2.039\times 10^{-13}\mbox{year}^{-1} $ for the
model in Case (i). As for the model in Case (ii), we set
$|\dot{G}/G|_{present} =2.042\times 10^{-13}\mbox{year}^{-1} $, resulting
in $|{\Delta G}/{G}| _{nucleosynthesis}\simeq 200\%$. Both values for
$|\dot{G}/G|_{present}$ clearly satisfy constraint {\bf2}).

For the above two cases, we calculate the evolution of the density
perturbations of dust $\Delta^E$ as the scale of the universe grows
$10^3$ times to today. We assume the wave number of the perturbations
$n=30$, which means today's ratio of physical wavelength of
perturbations to the Hubble horizon scale $\lambda_0/\lambda_ {H,0}=1/30$,
corresponding to large scale structure on scales of $\lambda_0
\simeq  100\mbox{Mpc}$. We normalize the initial condition for
$\Delta^E$($=\delta\mu/\mu$) to unity at the starting time. Therefore, it
is natural to set the initial conditions for
${\cal{Z}}^E$($=3\delta\hat{H}$) and ${\Phi}^E$($= \delta \varphi$) so as
to approximate roughly their background values $\hat{H}$ and
$\varphi$, respectively. We set
${\cal{Z}}^E\simeq-\hat{H}(=-32077\hat{H}_0)$ and
${\Phi}^E\simeq\varphi(=0.1903\sqrt{3}/\kappa)$ for Case (i), and
${\cal{Z}}^E\simeq-\hat{H}(=-51692\hat{H}_0)$ and
${\Phi}^E\simeq\varphi(=1.074\sqrt{3}/\kappa)$ for Case (ii) at the starting
time\footnote{We include the minus sign to pick up the growing
mode.}.

The evolution of $\Delta^E$ is shown in Fig. 1. The solid  line
represents  Case (i), while the  dashed line represents Case (ii).
Although $\Delta^E$ with all the constraints grows in almost the same
way as in  the Einstein de-Sitter case, that without {\bf 3}) is
greatly enhanced by a factor $\sim 6\times 10^2$. To see why, consider the
time evolution of gravitational ``constant''
${G}$ and its change rate $\dot{G}$, as shown in Fig. 2 and Fig. 3,
respectively. Corresponding to the enhancement in the growth rate of
$\Delta^E$, we see a great deviation of $\dot{G}$ from $\dot{G}|_{present}$
in Case (ii). On the other hand, the variations of gravitational constant
itself are within about several percent in both cases. This means that it
is not the variation of gravitational constant itself but its  change rate
that is the key to the enhancement of the growth rate of perturbations.

We emphasize that the enhancement of $\Delta^E$ shown above is a generic
feature for a wide range of parameters, such as $\Omega_{dust,0}$ or
$|{\dot{G}}/{G}| _{present}$. In order to see that the qualitative
behavior is the same for the case of
$\Omega_{dust,0}\ne 1.0$, consider Fig. 4, in which $\Omega_{dust,
0}=0.1$. We set
$|\dot{G}/G|_{present} =2.747\times 10^{-14}\mbox{year}^{-1}
$ for Case (i), $|\dot{G}/G|_{present} =2.889
\times 10^{-14}\mbox{year}^{-1} $ for Case
(ii), and set at the starting time ${\cal{Z}}^E=-19138\hat{H}_0$,
${\Phi}^E=1.186\sqrt{3}/\kappa$ and
${\cal{Z}}^E=-10023\hat{H}_0$,
${\Phi}^E=0.1228\sqrt{3}/\kappa$ for each case. The
enhancement is slightly reduced because of the lower density of matter.
As for the values of $|{\dot{G}}/{G}| _{present}$, as we mentioned above,
 constraint {\bf3}) strictly confines them. In order to satisfy
constraint {\bf3}) in the BD theory, $|{\dot{G}}/{G}| _{present}$ must be
almost exactly
$|\dot{G}/G|_{present} =2.039\times 10^{-14}\mbox{year}^{-1}$ for
$\Omega_{dust,0}= 1.0$ and
$|\dot{G}/G|_{present} =2.747\times 10^{-14}\mbox{year}^{-1}$ for
$\Omega_{dust,0}=0.1$. The other values
correspond to large deviations $|{\Delta G}/{G}| _{nucleosynthesis}$, and
$\Delta^E$ is greatly enhanced in most of such cases.

%
%
%
%
\section{Concluding Remarks}\label{sec5}\eqnum{0}
In this article we have explicitly presented gauge-invariant cosmological
perturbation equations in generalized Einstein theories which are
characterized as equations (\ref{3.2.1}) and (\ref{ap70}) with a perfect
fluid. Although such equations have been of interest to resolve
 the galaxy formation problem, the complex structure of the system has
made the formulation difficult. Therefore, we have first simplified the
fundamental equations through conformal transformation. This process
reduces the equations to the same form as in the Einstein theory
which allows us to apply many well-developed techniques to their solution.
This considerable simplification is important not only for this problem,
but for several others as well. Then we have
applied the covariant approach to derive the perturbation equations. This
approach makes the formulation straightforward, and the intuitive meanings
of the naturally introduced gauge-invariant variables are easy to obtain.
Now that its correspondence to the conventional Bardeen's approach is well
understood, this approach is becoming established and will be applied in
various situations to formulate relativistic perturbation theories.

As a simple example, we have applied the derived equations to
perturbations in Brans-Dicke cosmology. We have presented an analytic
solution for a specific background model, and shown an enhancement in
the growth rate of perturbations. We have also analyzed perturbations
in more generic models numerically. The enhancement of density
perturbations occurs throughout the range of the initial conditions or
parameters, so we have evaluated them assuming the observational
constraints. Although the three constraints we consider here are too strict
to permit a large enhancement  in density perturbations, if
we take into account only today's direct observations, we have found
 a growth rate high enough to account for galaxy formation.

 Our calculations imply that it is not the variation of effective
gravitational constant itself but its change rate that is the key to the
acceleration of the perturbation evolution. That means that if only our
universe experienced an  epoch in which the change rate of gravitational
constant was high, even if the variation itself was small, then
density perturbations could have grown fast enough to form the structure
observed today, and that other complex processes would be wholly
unnecessary. Hence the structure formation mechanism in  generalized
Einstein theories is very attractive. It may also be difficult  to rule out
completely such a possibility from observations. If GETs correctly describe
nature, they may predict enhanced growth rates for density perturbations.
We are now engaged in further studies of theories with non-minimally
coupled scalar fields and perfect fluid matter, to which we plan to apply
the results presented here.

\vspace{.5 cm}
\noindent
{\bf Acknowledgment}

We acknowledge L. D. Gunnarsen for his critical reading of our article.
This work was supported partially by the Grant-in-Aid for Scientific
Research  Fund of the Ministry of Education, Science and Culture  (No.
04640312 and No. 0521801), by a Waseda University Grant for Special
Research Projects.

\vspace{2cm}
\noindent {\large \bf Appendices}\vspace*{-0.5cm}\\

%
%
%
%
\appendix
\section{Formulation for the Most General Type of GETs}\eqnum{0}
In this appendix we consider the most general action for GETs,
i.e.,
\begin{equation}
S=\int d^4x \sqrt{-g}\left[ F(\phi,R)-\frac{\epsilon
(\phi)}2 (\nabla
\phi)^2+L_m\right],
\label{ap1}
\end{equation}
where $F(\phi, R)$ is an arbitrary function of a scalar
field $\phi$ and of a scalar curvature $R$. Here, we do not include Case
(A) (where $F$ is a linear function of $R$) because it has been discussed in
the text. Assuming a perfect fluid as (\ref{perfect}), the basic
equations are\cite{Maeda}
\bea
&&G_{ab}=\left( \frac{\partial F}{\partial R}\right)^{-1} \left\{
\frac{\epsilon(\phi)}2 \left[{\nabla}_a \phi {\nabla}_b\phi
-\frac12g_{ab}({\nabla}\phi)^2 \right] +\frac12
g_{ab}\left(F-\frac{\partial F}{\partial R}R \right) \right. \nonumber \\
&&\hspace{2.5em}+\left. \left[ {\nabla}_a {\nabla}_b \left( \frac{\partial
F}{\partial R}\right)-\square \left( \frac{\partial F}{\partial
R}\right)g_{ab}\right]  +\frac12 T_{ab}^{M}\right\},
\label{ap2}\\
&&\epsilon(\phi)\square\phi+\frac12
\frac{d\epsilon(\phi)}{d\phi}(\nabla\phi)^2 +\frac{\partial F}{\partial
\phi}=0. \label{ap3}
\ena
Although these involve higher-derivative terms of the
metric, the conformal transformation  (\ref{3.3.0}) with
\be
\omega=\frac12 \ln \left(2\kappa^2 \left|\frac{\partial F}{\partial
R}\right| \right),
\en
eliminates those  higher-derivatives, and the introduction of a new
``scalar'' field
\be
\kappa\varphi \equiv \sqrt{6}\omega=\frac{\sqrt{6}}2 \ln \left(2\kappa^2
\left|\frac{\partial F}{\partial R}\right| \right),\label{ap4}
\en
enables us to write down the equations in the Einstein form:
\begin{eqnarray}
&&{\hat{G}}_{ab} = {\kappa}^2
\left\{{\hat{\nabla}}_a\varphi {\hat
{\nabla}}_b\varphi-\frac12{\hat{g}}_{ab}(\hat{\nabla}\varphi)^2
\right\}+{\kappa}^2 E(\phi, \varphi) \left\{{\hat{\nabla}}_a\phi {\hat
{\nabla}}_b\phi-\frac12{\hat{g}}_{ab}(\hat{\nabla}\phi)^2  \right\}
\nonumber \\
&&\hspace{2.5em}-{\kappa}^2U(\phi, \varphi){\hat{g}}_{ab}+
{\kappa}^2N(\varphi) \left\{{\mu}
\hat{u}_a^{M} \hat{u}_b^{M}+p \hat{h}_{ab}^ {M}\right\} ,\label{a5}\\
&&\epsilon(\phi)\stackrel{\hat{}}{\square }
\phi+\frac12\frac{d\epsilon}{d\phi}(\hat{\nabla}\phi)^2-\kappa\epsilon(\phi)
\frac{\sqrt{6}}{3}{\hat{\nabla}}_a\varphi{\hat{\nabla}}^a\phi+\exp
\left[-\frac{\sqrt{6}}{3}\kappa \varphi \right]\frac{\partial F}{\partial
\phi}=0 ,
\label{a6}
\end{eqnarray} where
\begin{eqnarray}
E(\phi,\varphi)&=&(\mbox{sign})\epsilon(\phi)\exp\left[-\frac{\sqrt{6}}{3}
\kappa
\varphi \right],\qquad(\mbox{sign})=\frac{\partial F}{\partial
R}\left/\left| \frac{\partial F}{\partial R}\right|\right.,\\
 N(\varphi)&=&(\mbox{sign})\exp\left[-\frac{2}{3}\sqrt{6}\kappa \varphi
\right],\\
 U(\phi, \varphi)&=&N(\varphi)\left[\frac{\partial F}{\partial
R}R-F\right],\\
{\hat{u}}_a^{M} &=& e^{\omega}u_a^{M} ,\qquad {\hat{u}}^a_{M} =
e^{-\omega}u^a_{M} ,\\
 {\hat{h}}_{a}^{Mb} &=& h_{a}^{Mb },\qquad {\hat{h}}_{ab}^{M} =
e^{2\omega} h_{ab} ^{M} ,\qquad {\hat{h}}^{ab}_{M} = e^{-2\omega}
h^{ab}_{M} .
\end{eqnarray}
The (sign) equals $1$ for the present universe, and for a generic spacetime
there appears a singularity at ${\partial F}/{\partial R}=0$ beyond
which the (sign) turns to be $-1$. Then we should usually assume
$\mbox{(sign)}=1$ although we leave it in this article.
 The equation for $\varphi $ is obtained by taking the trace of
equation (\ref{ap2}) and using relation (\ref{ap4}):
\bea
\stackrel{\hat{}}{\square } \varphi+\frac{\sqrt{6}}{6}\kappa E(\phi,
\varphi)(\hat{\nabla}\phi)^2 +\frac{\sqrt{6}}{3} \kappa N(\varphi)
\left\{\frac{\partial F}{\partial R}R-2F-\frac12(-\mu+3p)\right\}=0 .
\label{a7}
\end{eqnarray}
In these equations, $R$ and $F$ are functions of
$\phi$ and $\varphi$ i.e., $R=R(\phi, \varphi)$ and $F=F(\phi,
\varphi)\equiv F(\phi, R(\phi, \varphi))$, which are defined implicitly
by the relation
\bea
\frac{\partial F}{\partial R}(\phi, R)&=&\mbox{a given function of  $\phi$
and $R$}\nonumber \\
&=&(\mbox{sign})\frac1{2\kappa^2}\exp\left[\frac{2}{\sqrt{6}}\kappa
\varphi\right].
\ena
The gravitational field equation (\ref{a5}) corresponds to the Einstein
one with two scalar fields and a perfect fluid interacting with each
other through
$E(\phi, \varphi)$, $U(\phi, \varphi)$ and $N(\varphi)$. The energy-momentum
tensor in  equation (\ref{a5}) is expressed in a fluid form like
equation (\ref{3.4.1}) as
\begin{eqnarray}
T_{ab}&=&\mu_* u_a^{O} u_b^{O} + p_*
h^{O}_{ab}+q_{*(a}u_{b)}^{O} + {\pi}_{*ab} ,
 \\
\mu_* &\equiv& \frac12 {\dot{\varphi}}^2+\frac12 E(\phi,
\varphi){\dot{\phi}}^2+U(\phi, \varphi)+N(\varphi){\mu} ,\\
p_* &\equiv&\frac12 {\dot{\varphi}}^2+\frac12 E(\phi,
\varphi){\dot{\phi}}^2-U(\phi, \varphi)+N(\varphi)p ,\\
q_{*a} &\equiv&-{\dot{\varphi}}^{(3)}{\nabla}_a\varphi-E(\phi,
\varphi){\dot{\phi}}^{(3)}{\nabla}_a\phi+N(\varphi)({\mu}+ p)V_a^{M} ,\\
{\pi}_{*ab} &\equiv& 0 ,\\
V_a^{M}&\equiv&u_a^{M}-u_a^{O},
\end{eqnarray}
to first order. Hence, the derivation of the GI
perturbation equations to  follow is straightforward, and proceeds as
described in
\S 3

Particularly in the case of $F=L(R)$ and $\epsilon(\phi)=0$ (Case (B)),
namely,  when the action is given by
\begin{equation} S=\int d^4x \sqrt{-g}\left[ L(R)+L_m\right],
\label{ap70}
\end{equation}
the basic equations become
\begin{eqnarray}
{\hat{G}}_{ab} &=& {\kappa}^2
\left\{{\hat{\nabla}}_a\varphi {\hat
{\nabla}}_b\varphi-\frac12{\hat{g}}_{ab}(\hat{\nabla}\varphi)^2 -U
(\varphi){\hat{g}}_{ab} \right\}\nonumber \\
&&+{\kappa}^2N(\varphi)\left\{{\mu}
\hat{u}_a^{M} \hat{u}_b^{M}+p \hat{h}_{ab}^ {M}\right\}  \label{LG}\\
\stackrel{\hat{}}{\square } \varphi &-& \frac{dU(\varphi)}{d\varphi}+
{\kappa}^2 M(\varphi)\{{\mu}-3 p\}=0 ,
\label{Lphi}
\end{eqnarray}
with
\bea
N(\varphi)&=&(\mbox{sign})\exp\left[-\frac{2}{3}\sqrt{6}\kappa \varphi
\right],\qquad(\mbox{sign})=\frac{\partial L}{\partial R}\left/\left|
\frac{\partial L}{\partial R}\right|\right.,\label{ap83}\\
U(\varphi)&=&N(\varphi)\left[\frac{\partial L}{\partial
R}R-L\right],\label{ap84}\\
M(\varphi)&=&\frac{\sqrt{6}}{6\kappa}N(\varphi).\label{ap85}
\ena
Here,
$R=R(\varphi)$ and $L=L(\varphi)\equiv L(R(\varphi))$ through the relation
\bea
\frac{\partial L}{\partial R}(R)&=&\mbox{a given function of
$R$}\nonumber \\
&=&(\mbox{sign})\frac1{2\kappa^2}\exp\left[\frac{2}{\sqrt{6}}\kappa
\varphi\right].\label{ap86}
\ena
Equations (\ref{LG}) and (\ref{Lphi})
have exactly the same form as equations (\ref{3.3.1}) and (\ref{3.3.2}) in
\S
\ref{sec3}. Hence the perturbation equations are given by
equations (\ref{delta}), (\ref{phi}) and (\ref{z}).

Among the models in Case (B), we are mostly
interested in $R^2$ gravity from the astrophysical point of view. By
setting $L(R)=\frac1{2\kappa^2}(R+CR^2)$ with $\mbox{(sign)}=1$,
equations (\ref{ap83})$\sim$ (\ref{ap86}) become
\bea
N(\varphi)&=&\exp\left[-\frac{2}{3}\sqrt{6}\kappa \varphi
\right],\label{ap87}\\
M(\varphi)&=&\frac{\sqrt{6}}{6\kappa}N(\varphi).\label{ap89}\\
U(\varphi)&=&\frac{N(\varphi)}{8\kappa^2C}
\left\{ 1-\exp\left[\frac{\sqrt{6}}{3}\kappa \varphi
\right]\right\}^2 ,\label{ap88}\\
\kappa\varphi &=& \frac{\sqrt{6}}{2}\ln \left[\left|1+2CR\right|\right].
\ena
In \cite{Rsquare}, perturbations in the $R^2$ gravity were  discussed.
However,  they were concerned only with metric perturbations without matter
fluid ($L_m=0$), and their approach was not GI, although their gauge
(conformal-Newtonian gauge) gave physical modes.

%
%
%
%
\section{Perturbations in a Scalar Field Dominated Universe}\eqnum{0}
In this appendix, taking the limit $\mu$, $p \rightarrow 0$, we show our
perturbation equations reduce to the ones in a scalar field dominated
universe derived by  Bruni et al.\cite{scalar}. Since they assumed the
Einstein theory, we must set
\bea  f(\phi)&=&\frac{1}{2\kappa^2},\\
\epsilon(\phi)&=&1,
\ena
in the action (\ref{3.2.1}), which immediately gives
\begin{eqnarray}
N(\varphi)&=&1,\\  M(\varphi)&=&0,\\
U(\varphi)&=&V(\phi). \label{relpot}
\end{eqnarray}
with $\varphi =\phi$. Besides, as they used center of mass frame $u_a^{E}$
, we can use equations (\ref{3.5.1})$\sim$(\ref{3.5.4}) as the basic
equations.

In a scalar field dominated universe, matter fluid perturbations are
absent, so we need consider only two independent GI perturbations. Bruni et
al. adopted the total energy density perturbation
$\Delta_*^E  =({a^2}/{{\mu}_*}){}^{(3)}{\nabla}_E^2{\mu}_*$ and the
curvature perturbation
$C^E = a^2{}^{(3)}{\nabla}_E^2\left[{^{(3)}}R\right]$ for them. They are
expressed in terms of our GI perturbation variables as
\bea
\Delta_*^E &=& \frac{\dot{\phi}^2}{\mu_*}\left(\frac{\dot{\phi}\dot{\Phi}^E
}{\mu}-\frac{V'{\Phi}^E }{\mu}\right),\label{reldelta}\\ C^E &=&
-4a^2H{\cal{Z}}^E+2\kappa^2a^2\mu_* \Delta_*^E.\label{relc}
\ena

We can recover their basic equations in the limit of $\mu$, $p \rightarrow
0$ in our equations. The background equations
(\ref{3.31})$\sim$(\ref{3.34}) become
\bea
&& 3\dot {H}+3H^2+{\kappa}^2(\dot{\phi}^2-V(\phi)) =0 ,\\
&& 3H^2+\frac{3k}{a^2}={\kappa}^2\mu_* ,\\
&&\ddot{\phi}+3H\dot{\phi} +V'(\phi)=0 ,\label{ddotphi}
\ena
where $\mu_* = \frac12 {\dot{\phi}}^2+V(\phi)$ and equation (\ref{ddotphi})
is equivalent to
\be
\dot{\mu}_*=-3H\dot{\phi}^2.
\en
The condition that total energy flux vanishes in the center of mass frame
implies in this limit
\be  u_a^{E} \rightarrow -\frac1{\dot{\phi}}\nabla_a \phi,
\en
which gives ${\Phi}^E \rightarrow 0$, so that $\Delta_*^E$ has a finite
value even in the limit of $\mu \rightarrow 0$.

In this limit, perturbation equation (\ref{delta}) is trivial. Equations
(\ref{z}) and (\ref{phi}) give respectively
\bea
&&\dot{\cal{Z}}^E=-2H{\cal{Z}}^E-\frac{\mu_*}{2\dot{\phi}^2}
\left(\kappa^2 \dot{\phi}^2 -\frac{2n^2}{a^2} +\frac{6k}{a^2}\right)
\Delta_*^E, \label{zlim}\\
&&\frac{\ddot{\Phi}^E}{\mu}=\left[3H+\frac{4V'}{\dot{\phi}}\right]
\frac{\dot{\Phi}^E}{\mu}+\left[-\frac{2V'^2}{\dot{\phi}^2}+V''\right]
\frac{{\Phi}^E}{\mu}-\frac1{\dot{\phi}}{\cal{Z}}^E.\label{philim}
\ena
By using the relation (\ref{reldelta}) and its derivatives, we can
eliminate ${\ddot{\Phi}^E}/{\mu}$, ${\dot{\Phi}^E}/{\mu}$ and
${{\Phi}^E}/{\mu}$ from equation (\ref{philim}) as
\be
\dot{\Delta}_*^E=\left[-3H+\frac{3H\dot{\phi}^2}{\mu_*}-\frac{\kappa^2
\dot{\phi}^2}{2H}\right] {\Delta}_*^E+
\frac{\dot{\phi}^2}{4a^2H\mu_*}C^E.
\en
Rewriting $3H$ for $\theta$ and using the notation $\gamma \equiv
\dot{\phi}^2/\mu_*$, this becomes
\be
\dot{\Delta}_*^E=\frac{3\gamma}{4a^2\theta}C^E+\left[(\gamma-1)\theta
-\frac{3\kappa^2\mu_*\gamma}{2\theta}\right]{\Delta}_*^E,\label{55}
\en
which corresponds to equation (55) in \cite{scalar}. Moreover we can
eliminate
${\cal{Z}}^E$ from equation (\ref{zlim}) by using equation (\ref{relc}) as
\bea
\dot{C}^E&-&2\kappa^2a^2\mu_*\dot{\Delta}_*^E \nonumber \\
&=&-HC^E-\frac{\kappa^2}{3H}(\dot{\phi}^2-V)C^E \nonumber \\
&&+\left[8\kappa^2a^2H\mu_*+\frac{2\kappa^4a^2\mu_*}{3H}(\dot{\phi}^2-V)
-6\kappa^2a^2H\dot{\phi}^2+2(-2n^2+6k)\frac{H\mu_*}{\dot{\phi}^2}
\right]{\Delta}_*^E.\nonumber\\
\ena
Replacing $\dot{\Delta}_*^E$ by equation (\ref{55}) and by using
equation (\ref{harmo}), this becomes
\be
\dot{C}^E=\frac{3k}{a^2\theta}C^E+\frac{4a^2\theta}{3\gamma}
{}^{(3)}{\nabla}^2{\Delta}_*^E+4k\left(\frac{\theta}{\gamma}
-\frac{3\kappa^2\mu_*}{2\theta}\right){\Delta}_*^E,\label{56}
\en
which corresponds to (56) in \cite{scalar}. The equations (\ref{55}) and
(\ref{56}) are the basic equations for GI perturbation variables
${\Delta}_*^E$ and $C^E$ given in \cite{scalar}.

%
%
%
%
%
\section{Analytic Solution in a BD Cosmology}\eqnum{0}
The solution (\ref{spephi})$\sim$(\ref{spemu}) of the flat dust BD universe
with the condition (\ref{special}) is described in terms of conformal
variables as
\bea
\frac{\hat{t}}{t_0}&=&\frac{4+3\omega}{5+3\omega}
\left(\frac{t}{t_0}\right)^{\frac{5+3\omega}{4+3\omega}}\label{tthat},\\
\varphi &=&\frac{\sqrt{2(3+2\omega)}}{\kappa(5+3\omega)}
\ln\left|\frac{5+3\omega}{4+3\omega}\frac{\hat{t}}{t_0}\right|,\\
\hat {a} &=&\left(\frac{5+3\omega}{4+3\omega}\frac{\hat{t}}{t_0}\right)
^{\frac{3+2\omega}{5+3\omega}},\\
\mu &=& \mu_0 \hat {a}^{-3}\left(\frac{\phi}{\phi_0}\right)^{-\frac32}
=\frac{2(3+2\omega)}{\kappa^2 t_0^2(4+3\omega)}
\left(\frac{5+3\omega}{4+3\omega}\frac{\hat{t}}{t_0}\right)
^{-\frac{6(1+\omega)}{5+3\omega}}.
\ena
Inserting this explicit background solution into equations
(\ref{reddelta})$\sim$ (\ref{redphi}), we present the perturbation
equations as
\be
\dot{\Delta}^E=-{\cal{Z}}^E+\frac{3(7+5\omega)\kappa}{\sqrt{2(3+2\omega)}
(4+3\omega)}\dot{\Phi}^E
 +\frac{\kappa  n^2 t_0}{\sqrt{2(3+2\omega)}}
\left(\frac{5+3\omega}{4+3\omega}\frac{\hat{t}}{t_0}\right)
^{-\frac{1+\omega}{5+3\omega}}\Phi^E,
\en
\begin{eqnarray}
\dot{\cal{Z}}^E=&-&\frac{(3+2\omega)(4+3\omega)}{(5+3\omega)^2}\frac1
{\hat{t}^2}{\Delta}^E\nonumber
\\ &-&\frac{2(3+2\omega)}{(5+3\omega)}\frac1{\hat{t}}{\cal{Z}}^E\nonumber\\
&-&\left[\frac{\sqrt{2(3+2\omega)}}{2(4+3\omega)}
\frac{\kappa}{\hat{t}}-\frac{\kappa  n^2 t_0}{\sqrt{2(3+2\omega)}}
\left(\frac{5+3\omega}{4+3\omega}\frac{\hat{t}}{t_0}\right)
^{-\frac{1+\omega}{5+3\omega}}\right]\dot{\Phi}^E\nonumber \\
&+&2\sqrt{2(3+2\omega)}\frac{4+3\omega}{(5+3\omega)^2}
\frac{\kappa}{\hat{t}^2}\Phi ^E,
\end{eqnarray}
\bea
\ddot{\Phi}^E=&-&\frac{2(4+3\omega)}{5+3\omega}\frac1{\hat{t}}\dot{\Phi}^E
\nonumber\\ &-&
\left[\frac{4(4+3\omega)^2}{(5+3\omega)^3}\frac1{\hat{t}^2}
+n^2\frac{4+3\omega}{5+3\omega}
\left(\frac{5+3\omega}{4+3\omega}\frac{\hat{t}}{t_0}\right)
^{-\frac{2(3+2\omega)}{5+3\omega}}\right]\Phi^E\nonumber\\
&+&\sqrt{2(3+2\omega)}\frac{(4+3\omega)^2}{(5+3\omega)^3}
\frac1{\kappa\hat{t}^2}{\Delta}^E\nonumber
\\ &-&\sqrt{2(3+2\omega)}\frac{(4+3\omega)}{(5+3\omega)^2}
\frac1{\kappa\hat{t}}{\cal Z}^E.
\ena Then we can solve these equations analytically as
\bea {\Delta}^E=&&
c_1\left\{\left(\frac{\hat{t}}{t_0}\right)^{\frac{2(2+\omega)}{5+3\omega}}
+\frac{4(8+5\omega)}{(5+3\omega)^2}
\left(\frac{5+3\omega}{4+3\omega}\right)^{\frac{1+\omega}{5+3\omega}}
\frac1{{t_0}^2 n^2}\right\}\nonumber \\
&+&c_2\left(\frac{5+3\omega}{4+3\omega}\frac{\hat{t}}{t_0}\right)
^{-\frac{4+3\omega}{5+3\omega}}\nonumber \\ &+&c_3
\left(\frac{\hat{t}}{t_0}\right)^{-\frac{4+3\omega}{2(5+3\omega)}}
J_{\nu}(nT) +c_4
\left(\frac{\hat{t}}{t_0}\right)^{-\frac{4+3\omega}{2(5+3\omega)}}
J_{-\nu}(nT),
\ena
\begin{eqnarray}
{\cal{Z}}^E=&-&c_1\frac{(3+2\omega)(4+3\omega)}{(5+3\omega)^2}\frac1{t_0}
\left(\frac{\hat{t}}{t_0}\right)^{-\frac{1+\omega}{5+3\omega}}\nonumber\\
&+&c_2 t_0\left(\frac{5+3\omega}{4+3\omega}\frac{\hat{t}}{t_0}\right)
^{\frac{1}{5+3\omega}}\nonumber \\  &+&c_3\left\{-\frac
n4\left(\frac{5+3\omega}{4+3\omega}\right) ^{\frac{2+\omega}{5+3\omega}}
\left(\frac{\hat{t}}{t_0}\right)^{-\frac{10+7\omega}{2(5+3\omega)}}
J_{\nu+1}(nT)\right.\nonumber \\
&&\qquad\quad+\left.\frac{n^2t_0}{4}\left(\frac{5+3\omega}{4+3\omega}
\right) ^{-\frac{1+\omega}{5+3\omega}}
\left(\frac{\hat{t}}{t_0}\right)^{-\frac{6+5\omega}{2(5+3\omega)}}
J_{\nu}(nT)\right\}\nonumber\\   &+&c_4\left\{\frac
n4\left(\frac{5+3\omega}{4+3\omega}\right) ^{\frac{2+\omega}{5+3\omega}}
\left(\frac{\hat{t}}{t_0}\right)^{-\frac{10+7\omega}{2(5+3\omega)}}
J_{-\nu-1}(nT)\right.\nonumber\\
&&\qquad\quad+\left.\frac{n^2t_0}{4}\left(\frac{5+3\omega}{4+3\omega}
\right) ^{-\frac{1+\omega}{5+3\omega}}
\left(\frac{\hat{t}}{t_0}\right)^{-\frac{6+5\omega}{2(5+3\omega)}}
J_{-\nu}(nT)\right\}, \end{eqnarray}
\bea  {\Phi}^E=&&c_1\sqrt{2(3+2\omega)}\frac{(8+5\omega)}{(5+3\omega)^2}
\left(\frac{5+3\omega}{4+3\omega}\right)^{\frac{1+\omega}{5+3\omega}}
\frac1{\kappa n^2 t_0^2 } \nonumber \\
&+&\frac{\sqrt{2(3+2\omega)}}{4\kappa}\left\{ c_3
\left(\frac{\hat{t}}{t_0}\right)^{-\frac{4+3\omega}{2(5+3\omega)}}
J_{\nu}(nT) +c_4
\left(\frac{\hat{t}}{t_0}\right)^{-\frac{4+3\omega}{2(5+3\omega)}}
J_{-\nu}(nT)\right\},
\nonumber \\
\ena
where $c_1$, $c_2$, $c_3$ and $c_4$ are arbitrary constants, and
\be
\nu\equiv \frac{4+3\omega}{2(2+\omega)},\qquad
T\equiv\frac{4+3\omega}{2+\omega}
\left(\frac{5+3\omega}{4+3\omega}\frac{\hat{t}}{t_0}\right)
^{\frac{2+\omega}{5+3\omega}}
\en

%
%
%
%
%

\vspace{2cm}
\noindent
{\large \bf Figure Captions}\vspace*{0.5cm}\\
Fig. 1:\\
Density perturbations in the BD theory for the case of
$\Omega_{dust,0}=1.0$. The solid  line represents  Case
(i); all the observational constraints {\bf 1)}, {\bf 2)} and {\bf 3)} are
satisfied, while the dashed line corresponds to Case (ii); only  {\bf 1)}
and {\bf 2)} are satisfied (see text). The density perturbation  in the
Einstein theory with the same initial condition coincides with
(i).\vspace*{1cm}\\ Fig. 2:\\
Evolution of effective gravitational constant ($G$=1/$\phi$). Each
line corresponds to the line in Fig. 1. bearing the same
number.\vspace*{1cm} \\
Fig. 3:\\
Evolution of  change rate of effective gravitational constant. Each
line corresponds to the line in Fig. 1. bearing the same number,
and $H_0=100$km/sec/Mpc.\vspace*{1cm}\\
Fig. 4:\\
Density perturbations in the BD theory  for the case
of $\Omega_{dust,0}=0.1$. The solid  line represents  Case (i); all the
observational constraints {\bf 1)}, {\bf 2)} and {\bf 3)} are satisfied,
while the dashed line corresponds to Case (ii); only  {\bf 1)} and {\bf 2)}
are satisfied (see text). The density perturbation  in the Einstein theory
with the same initial condition coincides with (i).

\end{document}